# Quantization of Light Energy Directly from Classical Electromagnetic Theory in Vacuum*

Wei-Long SHE [†]

Institute for Lasers & Spectroscopy, Sun Yat-Sen (Zhongshan) University, Guangzhou, China

It is currently believed that light quantum or the quantization of light energy is beyond classical physics and the picture of wave-particle duality, which was criticized by Einstein but attracted a number of experimental researches, is necessary for the description of light. We show in this paper, however, that the quantization of light energy in vacuum, which is the same as that in quantum electrodynamics, can be derived directly from the classical electromagnetic theory through the consideration of statistics based on classical physics. Therefore, the quantization of energy is an intrinsic property of light as a classical electromagnetic wave and has no need of being related to particles.

Keywords: the nature of light, quantization of light energy, light quantum, classical electromagnetic theory, consideration of statistics

1. Introduction

   Physicists have been trying to understand the nature of light over three and a half centuries. Light behaves like a continuous electromagnetic wave when it travels, and shows another face of discrete energy (momentum) when it interacts with charged particles. The latter leads to the concept of light quantum proposed by Einstein in 1905 [1]. However, what are light quanta? are they particles? This question has long been disturbing physicists, including Einstein himself. He even said that the problem was sufficient to drive him to the madhouse [2]. He had exerted himself to search for the origin of light quanta for fifty years [3-5]. But, even after the establishment of quantum electrodynamics (QED), he insisted in 1951 that there was still no answer to the question. He wrote, "All the fifty years of conscious brooding have brought me no closer to the answer to the question: What are light quanta? ...." [5] In fact, QED has not yet answered clearly whether light quanta are particles, even though it is a generally accepted theory [6]. In the many years that have elapsed, we have only started to scratch the surface although there have been some progress of the research on the light quanta [7]; we are today in the same state of "learned ignorance" with respect to light, as was Einstein [6]. And the light quanta still remain elusive [6]. It is currently believed that light quantum or the quantization of light energy is beyond classical physics and the picture of wave-particle duality, which was criticized by Einstein [3] but attracted a number of experimental researches [8-11], is necessary for the description of light.

   The modern particulate picture of light (or radiation) is built, to some extent, on the fact that the energy (momentum) of radiation is discrete when it is emitted and absorbed. Nevertheless, one could find that the experimental spectrum of $\gamma$-ray emission reported by Ruby and Bolef [12] can be well interpreted in the frame of classical electromagnetic theory. Besides, a number of works, which applied classical or semi-classical theory to explain the blackbody radiation spectrum [13], photoelectric effect [14], spontaneous emission [15, 16], stimulated emission [16], Compton effect and Lamb shift [17], reminded us once and again that the picture of particles (not QED) for light might be redundant. On the other hand, we know that: a) unlike the description for electrons in quantum mechanics, there is no proper Hermetian operator that straightforwardly corresponds to position for light quanta in quantum optics [6, 7]; b) in solid, the energy of an ultrasonic wave (a real wave) with a classical Hamiltonian (corresponding to a big group of one-dimensional harmonic oscillators) similar to that of a light field is quantized [18]. The difference of the description between electrons and light quanta, the comparability of two Hamiltonians corresponding respectively to an ultrasonic wave and a light field, and the obvious conjunction between an ultrasonic wave and its quantized energy remind us, from another aspect, that light may be, in a sense, the analogue of an ultrasonic wave and that there may exists a logical relation between a classical wave and its energy quanta, not only ultrasonic wave and phonons but also light wave and light quanta. In this paper, we will show that the quantization of light energy in vacuum can be derived directly from classical electromagnetic theory through the consideration of statistics based on classical physics and without any assumption of quantization.

2. Theory

   For convenience, we consider an electromagnetic radiation (or a light field) in a cavity. In classical electrodynamics, the energy of the radiation component having an angular frequency $\omega$, can be expressed as [19]

$$H = \sum_n \left(1/2 \cdot p_n^2 + 1/2 \cdot \omega^2 q^2{}_n\right)$$

(1)

where $q_n$ and $p_n$ are generalized coordinate and momentum, respectively; $n$ denotes the vibration mode of the electromagnetic field with frequency $\omega$. It is seen that the electromagnetic field can be treated as a mechanical system that contains a large number of one-dimensional harmonic oscillators (ODHOs) without interaction with each other. In a classical statistics point of view, the generalized coordinate and momentum of these ODHOs can be depicted as a whole by using two continuous random variables, $q$ and $p$ in a two-dimension phase subspace.

Let us focus our attention on the energy of these ODHOs. It is known that any conclusion of an acceptable theory of physics should be able to be tested by experiment and any result of experiment is statistic. Therefore any theoretical result of physics to be tested by experiment should be, in principle, statistic. So it is our interest to search for the expectation value of the ODHO's energy, $E$, instead of the exact energy of each ODHO. Avoiding those implications in the postulates in quantum mechanics, discarding the traditional statistical theory built on the postulate of equal a priori probabilities and having resulted in the energy equipartition theorem then the ultraviolet catastrophe, we try a new way here. According to classical physics and the possibility theory, the $E$ can be given by integral $E = \int_{-\infty}^{\infty} 1/2 \cdot p^2 \rho(p) dp + \int_{-\infty}^{\infty} 1/2 \cdot \omega^2 q^2 \rho(q) dq \equiv T + V$ if $\rho(q)$ and $\rho(p)$, the density functions corresponding respectively to $q$ and $p$, have been found out, where $T$ and $V$ stand for the expectation values of the kinetic energy and potential one of the ODHO, respectively. It is convenient to consider $T$ and $V$ dividually to reach the $E$. We first consider $V$ and the correlative $\rho(q)$. Mathematically, we can always introduce such a complex-valued function $\psi(q)$ that makes $|\psi(q)|^2 = \rho(q) \geq 0$. The $\psi(q)$ should then satisfy the following universal conditions though the form of it is still unknown:

$$\int_{-\infty}^{\infty} |\psi(q)|^2 \, dq = 1,$$

(2)

$$\lim_{|q| \to \infty} \psi(q) = 0,$$

(3)

$$V = \int_{-\infty}^{\infty} 1/2 \cdot \omega^2 q^2 |\psi(q)|^2 \, dq \leq E < \infty,$$

(4)

These are all the sound conditions available for the statistical description of $q$. They are understandable ones in classical physics: Eq.(2) is the general condition for any density function; Eq.(3) ensures the amplitude of each ODHO being finite; and Eq.(4) is the finiteness condition of the expectation value of the ODHO's energy, which is a well known one in classical physics. One can see that those meeting the conditions make up of the whole of the function candidates used to construct $\rho(q)$ (or calculate $V$) and the number of them is infinite.

Let $\delta\psi(q)$ be the variation of $\psi(q)$, and then $\delta\psi(q)$ can be taken as an arbitrarily tiny perturbation of the statistical distribution of $q$ in physics. The tiny perturbation is ineluctable since the absolutely ideal and perfect statistical distribution of $q$ cannot be found in the nature and the really statistical distribution is always accompanied with some tiny fluctuation. If a function satisfies conditions (2)-(4), it can be used to construct a density function $|\psi(q)|^2$ as a candidate describing the statistical distribution of $q$; if $\delta\psi(q)$ appears, then $|\psi(q)|^2$ becomes $|\psi(q) + \delta\psi(q)|^2$; and if $\psi(q) + \delta\psi(q)$ also satisfies conditions (2)-(4), then the statistical distribution described by $|\psi(q)|^2$ will be relatively stable and is a real one, otherwise is unstable and cannot appear independently. We shall show that *the relatively stable statistical distributions of $q$ and $p$ give the quantized $E$*.

To discuss the stability of the statistical distributions of $q$, we should change (2)-(4) into a set of equivalent "equilibrium conditions"(as does d'Alembert principle in which a dynamics equation is change into an equivalent equilibrium condition) since we cannot discuss the stability generally from conditions (2)-(4) [(2) can be taken as an equilibrium condition, but (4) cannot]. According to Eq.(2), $\psi(q)$ can be expanded into Fourier integration:

$$\psi(q) = (\omega/2\pi)^{1/2} \int_{-\infty}^{\infty} F(L)\exp(-i\omega qL)dL. \tag{5}$$

Condition (3) requires that

$$\lim_{|q|\to\infty} \int_{-\infty}^{\infty} F(L)\exp(-i\omega qL)dL = 0. \tag{6}$$

From Eqs.(2) and (5) it is known that

$$\int_{-\infty}^{\infty} |F(L)|^2 \, dL = 1. \tag{7}$$

Using derivative Fourier transform and the Parseval theorem under condition (4), one can immediately deduce (note that $\psi(q)$ is a slowly increasing distribution)

$$\int_{-\infty}^{\infty} |\frac{dF(L)}{dL}|^2 \, dL = \int_{-\infty}^{\infty} \omega^2 q^2 \, |\psi(q)|^2 \, dq < \infty. \tag{8}$$

Analogously, one can also show that (note that $F(L)$ is also a slowly increasing distribution)

$$\int_{-\infty}^{\infty} |\frac{d\psi(q)}{dq}|^2 \, dq = \int_{-\infty}^{\infty} \omega^2 L^2 \, |F(L)|^2 \, dL \tag{9}$$

if $\int_{-\infty}^{\infty} |d\psi(q)/dq|^2 \, dq < \infty$. The equation also keeps when $\int_{-\infty}^{\infty} |d\psi(q)/dq|^2 \, dq = \infty$ since in this case the supposition of $\int_{-\infty}^{\infty} \omega^2 L^2 \, |F(L)|^2 \, dL < \infty$ is contrary to the derivative Fourier transform and the Parseval theorem. So far, the equivalent "equilibrium conditions" (EECs) of (2)-(4) have been derived. They are Eqs.(2), (3), (6), (7), (8), and (9) [(2) is the sufficient condition of (5), (9) is the necessary condition of (5), while (6) and (7) are the necessary ones of (5) with (2) and (3), so (5) need not be involved in EECs; (8) is the sufficient and necessary condition of (4) with (2)]. One can see that if a function with variable $q$ is not the solution of EECs, it will not satisfy conditions (2), (3) and (4); and vice versa. Therefore, if a pair of functions $\psi(q)$ and $F(L)$ satisfy EECs but EECs cannot be hold anymore (the equilibrium is broken) after transformations $\psi(q) \to \psi(q) + \delta\psi(q)$ and $F(L) \to F(L) + \delta F(L)$ [note variation $\delta F(L)$ is arbitrary and independent of $\delta\psi(q)$], then the state of ODHOs described by $|\psi(q)|^2$ should not be able to appear independently, since in this case, $\psi(q) + \delta\psi(q)$ cannot satisfy (2)-(4) anymore, which means the statistical distribution described by this $|\psi(q)|^2$ is very unstable; conversely, if EECs are hold after the transformations, the state of ODHOs described by $|\psi(q)|^2$ will be relatively stable and is a real one.

To find out those relatively stable functions (corresponding to the relatively stable state of ODHOs) that satisfy EECs and then the conditions (2)-(4), we further turn to the following functional with two unknown functions $\psi(q)$ and $F(L)$ *:

$$I(\psi(q), F(L)) = 1/2\int_{-\infty}^{\infty} |\frac{d\psi(q)}{dq}|^2 \, dq - 1/2\int_{-\infty}^{\infty} \omega^2 L^2 \, |F(L)|^2 \, dL, \tag{10}$$

$$1/2\int_{-\infty}^{\infty} |\frac{dF(L)}{dL}|^2 \, dL - 1/2\int_{-\infty}^{\infty} \omega^2 q^2 \, |\psi(q)|^2 \, dq = 0, \tag{11}$$

$$\int_{-\infty}^{\infty} |F(L)|^2 \, dL - 1 = 0, \tag{12}$$

---

* The case described here is quite similar to that: since x=0 and y=0 (corresponding to $\psi_n(q)$ and $F_n(L)$ given in the following text) is the solution of equation $x^2 - y^2 = 0$ (corresponding to EECs) and it is at the saddle point of function $f(x, y) = x^2 - y^2$ (corresponding to the functional), the solution is relatively stable. Note: $x$ [corresponding to $\psi(q)$] is independent of $y$ [corresponding to $F(L)$] in function $f(x, y) = x^2 - y^2$ while it is not independent of $y$ in equation $x^2 - y^2 = 0$.

$$\int_{-\infty}^{\infty} |\psi(q)|^2 \, dq - 1 = 0, \tag{13}$$

$$\lim_{|q|\to\infty} \psi(q) = 0, \tag{14}$$

$$\lim_{|q|\to\infty} \int_{-\infty}^{\infty} F(L)\exp(-i\omega qL)dL = 0. \tag{15}$$

The functional is essentially different from that introduced by Schrödinger [20]. The solutions of the Euler-Lagrange's equation of Schrödinger's functional (Schrödinger's equation) cannot satisfy classical Hamilton-Jacobi equation, so Schrödinger's equation is only a hypothesis there; while the solutions of the Euler-Lagrange's equations of our functional [Eqs.(16) and (17), see below] satisfy classical EECs, so we have a logical route from classical conditions (2)-(4) to the quantization of the ODHO's energy.

It is evident that all the functions meeting EECs are involved in the admissible functions of this functional. So it is possible to use this functional to discuss whether $\psi(q)$ meets conditions (2)-(4) stably. By means of variational method, one can easily get the Euler-Lagrange's equations that determine the stationary functions from $\delta I = 0$:

$$-\frac{1}{2}\frac{d^2\psi(q)}{dq^2} + (-\lambda_1)\frac{\omega^2 q^2}{2}\psi(q) + \lambda_3\psi(q) = 0, \tag{16}$$

and

$$-\frac{\lambda_1}{2}\frac{d^2 F(L)}{dL^2} - \frac{\omega^2 L^2}{2}F(L) + \lambda_2 F(L) = 0, \tag{17}$$

where $\lambda_i$ ($i = 1, 2, 3$) denote three Lagrange's multipliers corresponding to constraints (11), (12) and (13), respectively. These $\lambda_i$ should be real ones under the constraints of Eqs.(11)-(15), in particular, $\infty > -\lambda_1 > 0$. Let $\beta_1 = (-1/\lambda_1)^{1/2}$, $\gamma = \lambda_3/\lambda_1$, and then Eq.(16) becomes $-\frac{\beta_1^2}{2}\frac{d^2\psi(q)}{dq^2} + \frac{\omega^2 q^2}{2}\psi(q) - \gamma\psi(q) = 0$ from which we can get the normalized eigenfunctions $\psi_n(q)$ ($n = 0, 1, 2, 3, \ldots$) having eigenvalues

$$\gamma_n = (2n+1)\beta_1\omega/2, \tag{18}$$

where $\beta_1$ is a parameter. Obviously, Fourier transforms $F_n(L) = (\omega/2\pi)^{1/2}\int_{-\infty}^{\infty}\psi_n(q)\exp(i\omega qL)dq$ ($n = 0, 1, 2, 3, \ldots$) are just the solutions of Eq.(17) with conditions Eqs.(12) and (15).

It is easy to verify [please see the **Appendix**] that $I(\psi_n(q), F_n(L)) = 0$, i.e., $\psi_n(q)$ and $F_n(L)$ satisfy EECs [note: if $\psi_n(q)$ and $F_n(L)$ cannot satisfy EECs, our logical route interrupts at once]. Moreover, we can show that $\psi_n(q) + \delta\psi(q)$ and $F_n(L) + \delta F(L)$ also satisfy EECs very well. Except for these {$\psi_n(q)$, $F_n(L)$}, any functions $\psi_a(q)$ and $F_a(L)$ satisfying EECs will lead to $\Delta I = \delta I(\psi_a(q), F_a(L)) \neq 0$, i.e., EECs break down after transformations $\psi_a(q) \to \psi_a(q) + \delta\psi(q)$ and $F_a(L) \to F_a(L) + \delta F(L)$. This means that {$\psi_n(q)$}, and only these {$\psi_n(q)$} are the relatively stable functions satisfying conditions (2)-(4). In a physics point of view, what will appear independently are the statistical distributions and only the statistical distributions described by {$|\psi_n(q)|^2$}. Now, $V$ can be ciphered out by using these {$\psi_n(q)$} according to the possibility theory. They are $\int_{-\infty}^{\infty} 1/2\,\omega^2 q^2 |\psi_n(q)|^2 \, dq = \gamma_n/2 = (2n+1)\beta_1\omega/4$ ($n=0,1,2,3\ldots$). Analogically, we can get that $T = (2m+1)\beta_2\omega/4$ ($m=0,1,2,3\ldots$), where $\beta_2$ is also a parameter. Let $\chi_1 = \beta_1/(\beta_1+\beta_2)$ and $\chi_2 = \beta_2/(\beta_1+\beta_2)$ (obviously, $\chi_1 + \chi_2 = 1$), and then we have

$$V = (2n+1)(\beta_1+\beta_2)\chi_1\omega/4 \qquad (n=0,1,2,3\ldots), \tag{20}$$

$$T = (2m+1)(\beta_1+\beta_2)\chi_2\omega/4 \qquad (m=0,1,2,3\ldots), \tag{21}$$

$$E = (\beta_1+\beta_2)/4 \cdot [(2n+1)\chi_1 + (2m+1)\chi_2]\omega \qquad (m,n=0,1,2,3\ldots). \tag{22}$$

On the other hand, it can be easily found for an ODHO in classical mechanics that $q = (2a/\omega)^{1/2} \sin Q$ and $p = (2a\omega)^{1/2} \cos Q$, where $Q$ and $a$ represent the new generalized coordinate and momentum (two new random variables now), respectively; $a$ is independent of $Q$. For any statistical distribution of $Q$, we always have

$$V = <1/2 \cdot \omega^2 q^2> = <a> \omega \cdot <\sin^2 Q>, \quad (23)$$

$$T = <1/2 \cdot p^2> = <a> \omega \cdot <\cos^2 Q>, \quad (24)$$

where $< \bullet >$ represents expectation value. It is easy to deduce from Eqs.(23) and (24) that $E = T + V = <a> \omega$. Comparing with Eq.(22), we have for the relatively stable states that

$$<a> = (\beta_1 + \beta_2)/4 \cdot [(2n+1)\chi_1 + (2m+1)\chi_2] \quad (m,n=0,1,2,3\ldots), \quad (25)$$

Obviously, $<\sin^2 Q>$ and $<\cos^2 Q>$ are independent of $m$ and $n$. Let $m, n = 0$ and we get from Eq. (25) that $<a_0> = (\beta_1 + \beta_2)/4$. Putting it into Eqs. (23), (24) and then comparing with Eqs.(20) and (21) for $m, n = 0$, one can reach $<\sin^2 Q> = \chi_1$ and $<\cos^2 Q> = \chi_2$. Moreover one will find $m = n$ from Eqs 20), (21), (23) and (24), which leads to $E = (2n+1)\beta\omega/2$ ( $n$ = 0, 1, 2, 3, …), where $\beta = (\beta_1 + \beta_2)/2$, a parameter, remaining to be determined by experiment. One can see that, to obey this energy law, the light (or radiation) can change its energy only by getting or giving one or more discrete units of energy such that $\varepsilon = \beta\omega$ if its frequency $\omega$ remains unchanged, which is consistent with the widely known experiences, such as the experiences of photoelectric effect and multiphoton absorption. Using the relation $\varepsilon = \beta\omega$ to fit the experimental result of photoelectric effect, one can find that $\beta$ determined by experiment is just the reduced Planck constant $\hbar$ and $\beta\omega$ is just the energy of Einstein's light quantum. Up to this point, the quantization of light energy, which is the same as that in quantum electrodynamics, has been derived directly from classical electromagnetic theory without any assumption of quantization. This demonstrates that the quantization of energy is an intrinsic property of light as a classical electromagnetic wave and has no need of being related to particles.

### 3. Conclusion

In conclusion, we have presented a theory to show that the quantization of light energy in vacuum can be derived directly from classical electromagnetic theory through the consideration of statistics based on classical physics. The theory reveals that the quantization of energy is an intrinsic property of light as a classical electromagnetic wave and has no need of being related to particles; the energy of Einstein's light quantum is the minimum change of the energy of classical electromagnetic radiation if the frequency of the radiation remains unchanged, which suggests that light quanta have no need of being endued with a particulate picture and should be taken as a convenient nomenclature describing the discrete, exchangeable energy (momentum) of classical electromagnetic radiation. Besides, the theory shows a logical route from classical ODHO to quantized one, which is *essentially* different from Nelson's one (in Nelson's theory, a hypothesis that every particle of mass $m$ is subject to a Brownian motion with diffusion coefficient $\hbar/2m$ is involved)[21]. One would not be surprised at the above result if he knew that a large number of random phenomena in nature obey the Gaussian distribution predicted by the central limit theorem in the possibility theory and noted that $|\psi_0(q)|^2$ is just a Gaussian distribution.

**Acknowledgments:** The mathematics applied in this paper has been discussed with Professors Yonglong DAI, Donggao DENG and Xiping ZHU. Professors Chengguang BAO, Fan ZHONG, Hong LU, and Dr. Ming LAI have read the manuscript. The author thanks them for their comments and suggestions.

### Appendix

According to variational principle, constraints (11)-(13) and boundary conditions (14), (15) are always satisfied for the admissible functions whether variations $\delta\psi(q)$ and $\delta F(L)$ appear or not. So, it is easy to verify that [note: $F_n(L)$ is the Fourier transform of $\psi_n(q)$]

$$1/2\int_{-\infty}^{\infty}|\frac{d\psi_n(q)}{dq}|^2\,dq - 1/2\int_{-\infty}^{\infty}\omega^2 L^2\,|F_n(L)|^2\,dL = 0, \tag{A1}$$

$$1/2\int_{-\infty}^{\infty}|\frac{dF_n(L)}{dL}|^2\,dL - 1/2\int_{-\infty}^{\infty}\omega^2 q^2\,|\psi_n(q)|^2\,dq = 0, \tag{A2}$$

$$\int_{-\infty}^{\infty}|F_n(L)|^2\,dL - 1 = 0, \tag{A3}$$

$$\int_{-\infty}^{\infty}|\psi_n(q)|^2\,dq - 1 = 0, \tag{A4}$$

$$\lim_{|q|\to\infty}\psi_n(q) = 0, \tag{A5}$$

$$\lim_{|q|\to\infty}\int_{-\infty}^{\infty}F_n(L)\exp(-i\omega qL)dL = 0, \tag{A6}$$

i.e. $\psi_n(q)$ and $F_n(L)$ satisfy EECs (2), (3), (6), (7), (8), and (9). Moreover, it is known that

$$\begin{aligned}
\Delta I &= 1/2\int_{-\infty}^{\infty}|\frac{d[\psi(q)+\delta\psi(q)]}{dq}|^2\,dq - 1/2\int_{-\infty}^{\infty}\omega^2 L^2\,|F(L)+\delta F(L)|^2\,dL \\
&\quad - [1/2\int_{-\infty}^{\infty}|\frac{d\psi(q)}{dq}|^2\,dq - 1/2\int_{-\infty}^{\infty}\omega^2 L^2\,|F(L)|^2\,dL] \\
&= \delta[1/2\int_{-\infty}^{\infty}|\frac{d\psi(q)}{dq}|^2\,dq - 1/2\int_{-\infty}^{\infty}\omega^2 L^2\,|F(L)|^2\,dL] \\
&\quad + 1/2[\int_{-\infty}^{\infty}|\delta(\frac{d\psi(q)}{dq})|^2\,dq - \int_{-\infty}^{\infty}\omega^2 L^2\,|\delta F(L)|^2\,dL] \\
&= \delta I + \frac{1}{2}\delta^2 I,
\end{aligned} \tag{A7}$$

where

$$\delta I = \delta[1/2\int_{-\infty}^{\infty}|\frac{d\psi(q)}{dq}|^2\,dq - 1/2\int_{-\infty}^{\infty}\omega^2 L^2\,|F(L)|^2\,dL], \tag{A8}$$

and

$$\delta^2 I = \int_{-\infty}^{\infty}|\delta(\frac{d\psi(q)}{dq})|^2\,dq - \int_{-\infty}^{\infty}\omega^2 L^2\,|\delta F(L)|^2\,dL. \tag{A9}$$

Ignoring the high order variation (high order infinitesimal) $\delta^2 I$, and then we have $\Delta I = \delta I$. Obviously [since $1/2\int_{-\infty}^{\infty}|\frac{d\psi_n(q)}{dq}|^2\,dq - 1/2\int_{-\infty}^{\infty}\omega^2 L^2\,|F_n(L)|^2\,dL = 0$ and $\delta I(\psi_n(q),F_n(L)) = 0$],

$$\begin{aligned}
&1/2\int_{-\infty}^{\infty}|\frac{d[\psi_n(q)+\delta\psi(q)]}{dq}|^2\,dq - 1/2\int_{-\infty}^{\infty}\omega^2 L^2\,|F_n(L)+\delta F(L)|^2\,dL \\
&= 1/2\int_{-\infty}^{\infty}|\frac{d[\psi_n(q)+\delta\psi(q)]}{dq}|^2\,dq - 1/2\int_{-\infty}^{\infty}\omega^2 L^2\,|F_n(L)+\delta F(L)|^2\,dL, \\
&\quad - [1/2\int_{-\infty}^{\infty}|\frac{d\psi_n(q)}{dq}|^2\,dq - 1/2\int_{-\infty}^{\infty}\omega^2 L^2\,|F_n(L)|^2\,dL] \\
&= \Delta I = \delta I(\psi_n(q),F_n(L)) = 0
\end{aligned} \tag{A10}$$

$$1/2\int_{-\infty}^{\infty}|\frac{d[F_n(L)+\delta F(L)]}{dL}|^2\,dL - 1/2\int_{-\infty}^{\infty}\omega^2 q^2\,|\psi_n(q)+\delta\psi(q)|^2\,dq = 0, \tag{A11}$$

$$\int_{-\infty}^{\infty}|F_n(L)+\delta F(L)|^2\,dL - 1 = 0, \tag{A12}$$

$$\int_{-\infty}^{\infty}|\psi_n(q)+\delta\psi(q)|^2\,dq - 1 = 0, \tag{A13}$$

$$\lim_{|q|\to\infty}[\psi_n(q)+\delta\psi(q)]=0 \quad , \tag{A14}$$

$$\lim_{|q|\to\infty}\int_{-\infty}^{\infty}[F_n(L)+\delta F(L)]\exp(-i\omega qL)dL=0, \tag{A15}$$

i.e. $\psi_n(q)+\delta\psi(q)$ and $F_n(L)+\delta F(L)$ also satisfy EECs very well. {Note: According to variational principle, $\delta I=0$ with conditions (11)-(15) satisfied is equivalent to $\delta J=0$, where $J=I+\lambda_1\cdot[Left\ of\ Eq.(11)]+\lambda_2\cdot[Left\ of\ Eq.(12)]+\lambda_3\cdot[Left\ of\ Eq.(13)]$ }.Except for these { $\psi_n(q)$ and $F_n(L)$ }, any functions $\psi_a(q)$ and $F_a(L)$ meeting EECs and making $1/2\int_{-\infty}^{\infty}|\frac{d\psi_a(q)}{dq}|^2 dq - 1/2\int_{-\infty}^{\infty}\omega^2 L^2 |F_a(L)|^2 dL=0$ will lead to [since $\delta I(\psi_a(q),F_a(L))\neq 0$]

$$\begin{aligned}&1/2\int_{-\infty}^{\infty}|\frac{d[\psi_a(q)+\delta\psi(q)]}{dq}|^2 dq - 1/2\int_{-\infty}^{\infty}\omega^2 L^2 |F_a(L)+\delta F(L)|^2 dL\\ &=1/2\int_{-\infty}^{\infty}|\frac{d[\psi_a(q)+\delta\psi(q)]}{dq}|^2 dq - 1/2\int_{-\infty}^{\infty}\omega^2 L^2 |F_a(L)+\delta F(L)|^2 dL\\ &\quad -[1/2\int_{-\infty}^{\infty}|\frac{d\psi_a(q)}{dq}|^2 dq - 1/2\int_{-\infty}^{\infty}\omega^2 L^2 |F_a(L)|^2 dL]\\ &=\Delta I=\delta I(\psi_a(q),F_a(L))\neq 0\end{aligned} \tag{A16}$$

$$1/2\int_{-\infty}^{\infty}|\frac{d[F_a(L)+\delta F(L)]}{dL}|^2 dL - 1/2\int_{-\infty}^{\infty}\omega^2 q^2 |\psi_a(q)+\delta\psi(q)|^2 dq=0, \tag{A17}$$

$$\int_{-\infty}^{\infty}|F_a(L)+\delta F(L)|^2 dL-1=0, \tag{A18}$$

$$\int_{-\infty}^{\infty}|\psi_a(q)+\delta\psi(q)|^2 dq-1=0, \tag{A19}$$

$$\lim_{|q|\to\infty}[\psi_a(q)+\delta\psi(q)]=0 \tag{A20}$$

$$\lim_{|q|\to\infty}\int_{-\infty}^{\infty}[F_a(L)+\delta F(L)]\exp(-i\omega qL)dL=0 \tag{A21}$$

i.e. $\psi_a(q)+\delta\psi(q)$ and $F_a(L)+\delta F(L)$ cannot satisfy EECs any more.

Here is a numerical example which would be helpful for understanding the above theory. Let $\psi_a(q)=\sqrt{c}\psi_0(q)+\sqrt{d}\psi_1(q)$ [ $c+d=1$ ] and then we choose such a variation $\delta\psi(q)=\psi_a(q)\sum_{j=1}^{200}\rho_{ij}\sin(j\frac{\pi}{b}q)$ for $\psi_a(q)$ (note: $\delta\psi(q)$ is arbitrary), where { $\rho_{ij}$ }are some of random numbers obeying the uniform distribution in [-0.00005, 0.00005], and $b$ is a positive number large enough. Taking 80 sets of { $\rho_{ij}$ } [$i=1, 2, 3,…, 80$; for a fixed $i$, $j$ is changed from 1 to 200, and then a set of random numbers is gotten, for example, $i=5$, { $\rho_{ij}$ }=( $\rho_{51},\rho_{52},\rho_{53},...\rho_{5200}$ )] and calculating integration

$$w_i=\int_{-b}^{b}(\psi_a(q)+\delta\psi(q))^2 dq-1$$

for different $c$, we get the results shown as Fig.1, where c=1 and c=0 correspond respectively to $\int_{-b}^{b}(\psi_0(q)+\delta\psi(q))^2 dq-1$ and $\int_{-b}^{b}(\psi_1(q)+\delta\psi(q))^2 dq-1$; $w_i$ reflects the departure of the state with $\delta\psi(q)$ from the ideally statistical distribution. The cases of c=0.9. 0.99 and 0.999 are the same as those correspond to c=0.1, 0.01 and 0.001, respectively. One can see that the statistical distributions described by $|\psi_0(q)|^2$ and $|\psi_1(q)|^2$ are relatively stable while that described by $|\psi_a(q)|^2$ is unstable. It should be pointed out that under the constraint of admissible function the instability of $|\psi_a(q)|^2$ shown here is transferred to $I(\psi(q), F(L))$ in the above variational procedure.

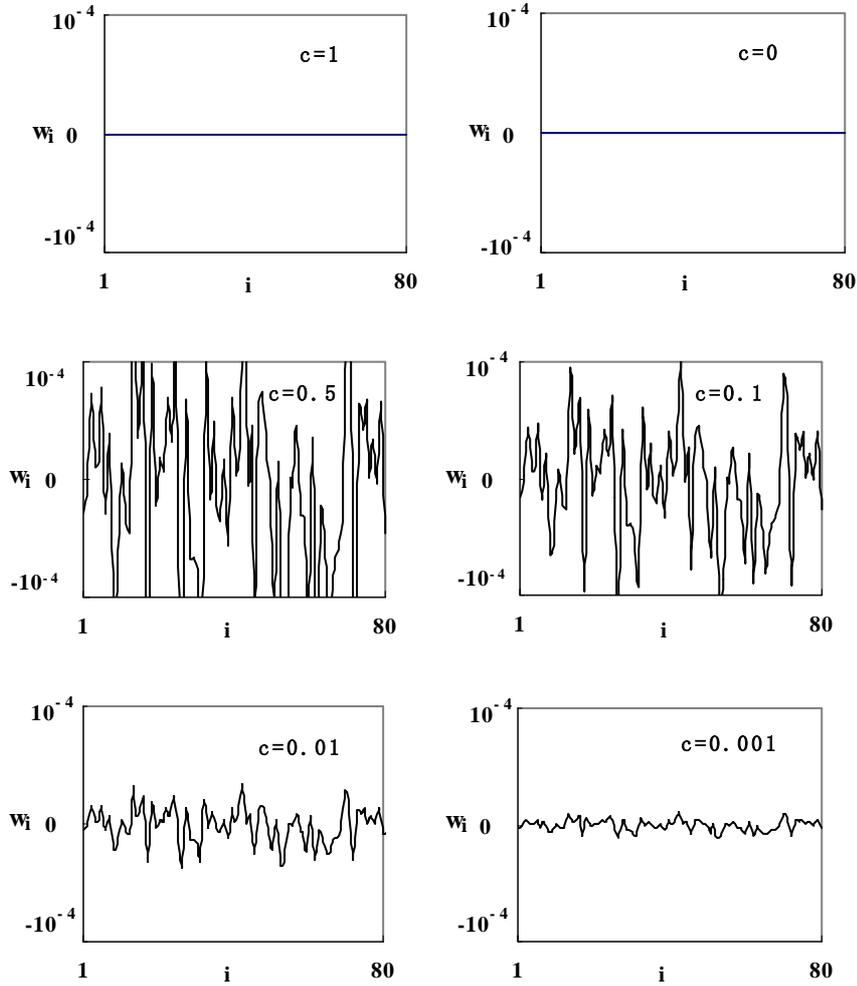

Fig.1 The stability of the statistical distributions described by $|\psi_0(q)|^2$, $|\psi_1(q)|^2$ and $|\psi_a(q)|^2$, where $\psi_0(q) = \sqrt{\alpha}/\pi^{1/4} \cdot \exp(-1/2 \cdot \alpha^2 q^2)$, $\psi_1(q) = \sqrt{2\alpha}/\pi^{1/4} \cdot \alpha q \exp(-1/2 \cdot \alpha^2 q^2)$, and $\psi_a(q) = \sqrt{c}\psi_0(q) + \sqrt{d}\psi_1(q)$ [$\alpha = 1, c + d = 1$]; $w_i = \int_{-b}^{b} (\psi_a(q) + \delta\psi(q))^2 dq - 1$, in which $\delta\psi(q) = \psi_a(q) \sum_{j=1}^{200} \rho_{ij} \sin(j\frac{\pi}{b} q)$, {$\rho_{ij}$} are some of random numbers obeying the uniform distribution in [-0.00005, 0.00005], and $b$ is a positive number large enough.